\documentstyle[12pt,iopconf]{article}
\def\mxth{\mathsurround=0pt }

\def\xversim#1#2{\lower2.pt\vbox{\baselineskip0pt \lineskip-.2pt
    \ialign{$\mxth#1\hfil##\hfil$\crcr#2\crcr\sim\crcr}}}
\def\ltsim{\mathrel{\mathpalette\xversim <}}
\def\gtsim{\mathrel{\mathpalette\xversim >}}

\begin{document}
\title
{\bf Recent advances in extended inflationary cosmology\footnote{Invited Talk
at the {\it Journ\'{e}es
Relativistes}, Amsterdam, May 14-16, 1992}}
\author{
  Paul J. Steinhardt\dag}
\affil{\dag\ Department of Physics,
University of Pennsylvania,
Philadelphia, PA  19104}

\beginabstract
Extended inflation is a promising new approach to implementing the
inflationary universe scenario.  This paper reviews recent advances including
a new, more robust
mechanism for ending extended inflation, a new prediction for  the
density fluctuation spectrum generated by extended inflation, and
the discovery that extended inflation can   produce gravitational
waves that can significantly add  to the cosmic background anisotropy.
\endabstract

\section{Introduction}
The inflationary universe hypothesis
is appealing because it offers elegant solutions
to longstanding cosmological problems (Guth 1981; Guth and Steinhardt 1989)
that plague the standard big bang picture.  However, implementing
the scenario in accordance with astrophysical constraints has required
an extreme and unnatural fine-tuning of parameters when
constructing detailed models.  Recently, new approaches
for implementing inflationary cosmology --- known generically as {\it extended
inflation} ---  have been developed which may remove the fine-tuning problem
(La and Steinhardt 1989).
The new approaches  also lead to important  new effects  on gravitation,
the gravitational wave background,
the cosmic microwave background
and the large-scale structure of the universe.

Several significant advances in extended inflationary cosmology
have been made within the last few months.
 First,
 two qualitatively distinct mechanisms have been identified
 for ending  inflationary expansion.  As will be described below,
    the  mechanism  invoked by
 earlier authors
  can
  produce big bubbles that unacceptably
  distort the cosmic microwave background unless
  parameters are carefully chosen.
  The  new mechanism  (Crittenden and Steinhardt 1992)
   easily
  avoids these problems and perhaps the fine-tuning problem.
  Second, it has been shown that
  the   spectrum of density fluctuations produced via this new mechanism
    is scale-free but not Harrison-Zel'dovich (Crittenden and
    Steinhardt 1992):
      $P(k) \propto k^n$, where $n$ lies
      between 0.5 and 1.0  (whereas, previously, it appeared that
      the range $1.0> n \gtsim 0.84$ was
      disallowed).
        The predictions are consistent with the
      recent COBE observations of cosmic microwave background (CMB) anisotropy.
      Finally, shortly after {\it Journ\'{e}es Relativistes 1992},
      it was discovered that extended inflation can generate
  large amplitude gravitational waves which add to,  perhaps even
  dominate, the  CMB anisotropy (Davis \etal 1992).
  This latest discovery will be
detailed in a  separate
article by  G. F. Smoot and the author  that can be found in
this volume.  Before describing the new developments, we will review
the qualitative differences among inflationary models.

Section~2 is a qualitative overview of earlier inflationary
models intended to point out the weakness and, thereby, to
explain the motivation for seeking new approaches.
 Section~3 is a descriptive introduction to extended inflation.
 Section~4 focuses on the \underline{amplitude} of the nearly
 scale-invariant spectrum of density
 fluctuations generated by extended inflation.
 This is a central issue since obtaining small-amplitude
 fluctuations has been the most difficult challenge for inflationary
 model-building.  The results depend sensitively on how inflation
 terminates. This is why the new mode of ending extending inflation
 described in Section~6 is so significant.  In Section~5 we
 point out a fascinating byproduct of extended inflation ---
 a cosmic remnant of bubble collisions in the form of
 a unique signature in the gravitational radiation spectrum.
 In Section~7, we return to the density fluctuation spectrum
 from extended inflation, but this time we focus on the
 \underline{scale-dependence}.  The traditional view is that
 inflation predicts a scale-invariant spectrum.  Here we show
 that there is a a somewhat wider range of possibilities, a
 conclusion that is especially important in light of the recent
 COBE DMR results.

\section{An overview of earlier inflationary models}

The goal of every inflationary model is to generate a brief period in which the
scale factor of the universe, $a(t)$, increases superluminally, $a(t)>t$.
If $a(t)$ grows by ${\rm e}^{60}$ or more during this period, the cosmological
horizon, flatness and monopole problems can be resolved.  In addition,
inflation generates  energy density fluctuations which may be seeds for
galaxy formation.

Designing a detailed microphysical
model that accomplishes all  of these goals has
proven to be extremely difficult.  One needs  a
  {\it grand entrance} into inflation: a robust mechanism that
  drives the universe into a false vacuum phase. The large,
  positive vacuum density  acts as an effective cosmological
  constant that triggers a period of  de Sitter expansion.
  Then, one needs a {\it graceful exit}: a mechanism to
  teminate the de Sitter expansion, reheat the universe to
  a high temperature, and restore Friedmann-Robertson-Walker
  expansion.
The plethora of models have been proposed in the past decade
can be broadly categorized according to their entrance and
graceful exit.
   See Figure 1.

 \begin{figure}
 \vspace*{6in}
 \caption{ }
 \end{figure}

 The original,  ``old inflation'' model (Guth 1981) relied on a simple
 and compelling mechanism for grand entrance: a strongly, first order
 phase transition.  The mechanism seems natural enough  because nearly
 every kind of unified theory predicts a whole sequence of phase
 transitions in the first instants after the big bang
 associated with
 spontaneous symmetry breaking.
 The transition is driven by a scalar field  (or set of fields)
 $\sigma$, known
 as the inflaton, whose expectation value changes as the universe
 transforms from the metastable (false) to stable (true)
 phase.  The evolution of
 the inflaton is determined by the finite-temperature effective
 potential for $\sigma$.

 At temperature well above the critical temperature for the transition,
 the metastable phase is entropically favored and, consequently, has
 the lowest free energy density.
 As the temperature decreases below the critical
 temperature, the free energy density of the stable phase falls
 below that of the metastable phase, but the universe remains
 trapped  in the false phase by an energy barrier that separates
 the two phases. That is,
 the universe supercools into the false vacuum phase.
 So, in a very natural way, the false vacuum  energy density
 comes to dominate the energy density of the universe, and
 inflationary expansion commences.

However, while the entrance is quite grand, the exit is not so graceful.
 The problem with  old
 inflation is that there is no
 way to escape the false vacuum and return to
  a hot, Friedmann-Robertson-Walker
 universe
 (Guth and Weinberg 1983).   The usual
 mechanism for completing strongly first order phase transitions is
  bubble nucleation, the tunneling of a finite region from the false
  vacuum
  phase through the energy barrier to the  true vacuum phase.  The
  bubbles spontaneously form, grow, and coalesce to complete the
  transition.  Here, though,  space between the forming bubbles
  expands superluminally, too fast  for the bubble walls to ever
  meet and complete the transition.  We will re-examine this
   failure in
  more detail when we discuss extended inflation.

``New inflation" models (Linde 1982; Albrecht and Steinhardt 1982)
 employ the same grand entrance as old inflation, but  introduce a
 new kind of graceful exit.   A special kind of phase transition is
 assumed in which the energy  barrier separating the metastable
 and stable phases disappears during supercooling.
 In condensed matter parlance, this corresponds to a ``spinodal"
 rather than ``nucleation-driven" first order transition. The entire
 universe is then free to evolve continuously
  (``slow-roll'') from the false to
 the true vacuum phase.  This process can be described as
 having $\sigma$ roll continuously along the effective
 potential from its false vacuum
 expectation value to its true vacuum expectation value.
  Near the end of the slow-roll, the effective potential steepens
  and $\sigma$ begins to oscillate rapidly about the true
  vacuum expectation value.  The oscillating field radiates (decays)
   itself and other particles, rapidly thermalize.  In this way,
   the false vacuum energy density is ultimately restored as
   thermal energy density.

 The key problem with new inflation is that
  the slow-roll requires
 extreme fine-tuning of parameters in the effective
 potential, especially to ensure an
 acceptable distribution of density fluctuations after the
transition (Bardeen \etal 1983; Guth and Pi 1982;
Hawking 1982; Starobinskii 1982).

``Chaotic inflation" models (Linde 1983) introduce a novel grand
entrance.  The phase transition is abandoned as a means of
driving the universe into the false vacuum phase.  Instead,
one
invokes chaotic initial conditions: perhaps quantum or
thermal fluctuations drive large regions of the universe to
 high energy density enough to trigger inflation.  How grand
 this entrance is depends upon one's view about
 how  generic such chaotic conditions are. However, the key
 point is that the graceful exit mechanism is identical to
 that of new inflation: slow-roll from the   false to
  the true vacuum phase.
 Consequently, {\it chaotic inflation requires exactly the same
 fine-tuning of parameters as new inflation},
 a point which seems to have been confused in the literature.

 Power-law inflation  (Lucchin and Matarrese 1985) is a variant of
 chaotic-inflation in which the  energy potential is designed so
 that the  slow-roll is not so slow:  instead of the universe
 expanding exponentially with time, it expands as a power-law
 $a(t) \propto t^{\alpha}$ with $\alpha > 1$. To achieve this,
 one replaces the  effective potentials of chaotic inflation,
 which are polynomial in $\sigma$, with a potential that
 is exponential in $\sigma$.  Fine-tuning is still required
 to obtain a slow-roll with acceptable density fluctuations.
 Yet additional fine-tuning is required to modify the
 exponential potential to  have a true vacuum phase  about
 which $\sigma$ can oscillate for reheating.

 Given the thorny, technical problems with
 each of these approaches and  yet the
 increasing attractiveness of the inflationary hypothesis,
  a critical
  challenge for cosmologists is to find improved approaches.

\section{Extended inflation}

Extended inflation returns to Guth's
original, highly attractive,
``old'' inflation  concept:
  Extended inflation is also initiated when
  the universe is trapped in a false vacuum state by a large energy
  barrier during the course of a strongly first-order phase transition.
  The failure of old inflation was that the universe could
  never escape the false vacuum state because the rate for tunneling
  through the barrier remains small compared to the inflationary
  expansion rate (Guth and Weinberg 1983).
Extended inflation avoids the same failure
by introducing
mechanisms            so
that the tunneling rate eventually surpasses
the expansion rate and, hence, the transition to the true vacuum
can be completed.

The fact that first order cosmological phase transitions can
be completed was not known prior to extended inflation
and has two immediate and important implications.
First,
the cosmological ban on strongly first order phase transitions
is lifted.  Previously, the failure of old inflation was used to
argue that there could never have been a strongly first-order
phase transition since the big bang or else the universe would
have been trapped forever in an inflationary false phase (Linde 1977;
Guth and Weinberg 1980).
This was applied as a constraint on particle physics models.
With the discovery of new
mechanisms for successfully completing transitions
by bubble nucleation,                 strongly first
order phase transitions, whether inflationary or not, are
permissible, thereby opening the range of possible parameters.
Second, in an ordinary first order transition,
the fields driving the phase transition do not need to be weakly
coupled  as has been  required by prior inflation models
based on slow-roll.  Not
only does this relieve the fine-tuning of the effective
potential, but it eliminates any
problems with reheating after
inflation (Steinhardt and Turner 1984).

To explain how extended inflation succeeds, it is useful to review
how old inflation fails.  Success or failure depends on
the relation between the tunneling or bubble-nucleation rate, $\lambda$,
to  the expansion rate (Hubble parameter), $H$.
 The ratio is the dimensionless bubble nucleation parameter,
$\epsilon \equiv \lambda/
H^{4} $.  Guth and Weinberg (1983)
earlier showed that the false vacuum
can be percolated by true vacuum bubbles only if $\epsilon$ exceeds
a critical value, $\epsilon_{crit} \approx .02$.  In old inflation,
$\epsilon$ is time-independent since $\lambda$ (which depends on
the barrier shape) and $H$ (which depends on the false vacuum energy)
do not vary during inflation.  Two dismal       fates are possible:
(a) $\epsilon <  \epsilon_{crit}$, in which case the true vacuum
bubbles never percolate and the universe inflates forever; or,
(b) $\epsilon >
\epsilon_{crit}$, in which case the true vacuum percolates, but so
quickly that there is insufficient inflation to solve any cosmological
problems.

One method of avoiding the dismal      fates is to avoid bubble
nucleation altogether.   New inflation and chaotic inflation
utilize this approach.  For example, in new inflation, the energy
barrier disappears altogether as the universe supercools and the
universe evolves     slowly but continuously
 from the false to true vacuum phase (Linde 1982; Albrecht and
 Steinhardt 1982).

   Extended inflation models employ a different, almost obvious
   (in retrospect)
   alternative approach:  instead of allowing $\epsilon$ to remain
   time-independent, mechanisms are introduced so that $\epsilon$ is
   initially much less than $\epsilon_{crit}$ to achieve sufficient
   inflation, but then grows during inflation to a value $\epsilon >
   \epsilon_{crit}$ so that the phase transition can be completed.
   For typical (untuned) first order phase transitions, $\epsilon$
   is exponentially small at the onset of inflation, ${\cal O}(
    {\rm e}^{-40})$   or smaller.  Hence, an increase in $\epsilon$
    by 20 or more orders of magnitude is sought.

How can $\epsilon \equiv \lambda/H^{4}$ be made time-dependent?
Although there  are a variety of approaches
(Steinhardt 1990), perhaps the simplest
and most natural is to introduce a modification
of the  Einstein gravitational action.
Modifying Einstein gravity may seem to be a wildly radical
option to some, so  a few remarks are
in order:  Although
Einstein gravity has proven
to be a remarkably accurate description of gravity based on both
local and cosmological tests,
modified gravity theories are well-motivated.
Modifications of Einstein gravity are virtually
mandated by quantum mechanics.  Any attempt to unify particle physics
with gravity seems ultimately forced to include some modifications of
Einstein gravity.  If one begins with Einstein gravity plus particle
fields at tree level, quantum corrections alter the gravitational
interaction.     To be sure,       the resulting
theory is    not renormalizable and
the new interactions   are only finite if one introduces a
cutoff.  However, there is no reason to believe that pure Einsteinian
gravity will result when the ultimate, finite unified theory is
developed.  If one looks at the best candidates for unified theories
to date, supergravity
and superstring theories, one sees that modifications of Einstein
gravity are essential to the structure of the theory (e.g., the
interactions of dilaton and moduli fields).  Certainly, the
correct quantum gravity theory, whatever it is, must converge
to Einstein gravity at low energies.  But there is no reason
to expect the quantum corrections to be small at the very
high energies associated with inflation.  So, in that spirit,
it is natural to investigate if such modifications are useful
(or hurtful) to inflationary cosmology.

What one finds is that modifications to Einstein gravity
 can profoundly change inflation.
A key effect is that the effective  Newton's
constant $G$ can be made to  decrease
during inflation for an extremely broad
range of models.  Decreasing $G$  leads to  decreasing
Hubble parameter $H^2 = 8 \pi G \rho_F/3$
which, in turn, leads to increasing $\epsilon \equiv \lambda/
H^{4} $, just what was desired!

As a specific example, consider the Lagrangian density
\begin{equation}
{\cal L}_{G}
=  \frac{f(\phi)}{G} {\cal R}
- \frac{1}{2}(\partial_{\mu} \phi)^{2}
- \frac{1}{2}(\partial_{\mu} \sigma)^{2}
-V(\sigma) + {\cal L}_{other},
\end{equation}
where $f(\phi)$  is the non-minimal coupling between a scalar
field $\phi$ and the scalar curvature ${\it cal R}$.
Here we imagine that $\sigma$  remains as the order parameter
field for the inflationary phase transition and that $\phi$ is
a completely independent field.  As the universe supercools into
the false vacuum phase ($\sigma =0$, say), the false vacuum
energy density $V(\sigma)$ dominates the energy density and
  drives the universe towards superluminal expansion.  An
  added effect is that the scalar curvature becomes non-zero, which
  creates a back-reaction on the independent field $\phi$.  If
  $f(\phi)$ is positive and increasing, then the back-reaction
  forces $\phi$ to change such that $f(\phi)$ grows.
 Since the effective Newton's constant is
$G_{eff} = G/f(\phi)$,
 increasing $f(\phi)$ results in decreasing $G_{eff}$, decreasing
 $H$, and increasing $\epsilon$.
 The equation of motion for $\phi$ is:
 \begin{equation}   \label{eom}
 \ddot{\phi} + 3 H \dot{\phi} + \frac{1}{2} \frac{(1 + 6 f'') f'
 \dot{\phi}^2}{f + 3 f'^{2}} - \frac{ 2 f'
 V(\sigma)}{f  + 3 f'^{2}} = 0
 \end{equation}

There is great freedom in choosing $f(\phi)$, so long as it
increases during inflation.
   In the first
   extended inflation model (La and Steinhardt 1989),
    $f(\phi)$ was set equal to $\xi \phi^{2}$,
    resulting in
    a  model that is mathematical equivalent to a Brans-Dicke
    theory (Brans and Dicke 1961)
     with Brans-Dicke parameter, $b=1/8 \xi$.
     For pure Brans-Dicke theories in which $\phi$ continues to evolve
     to the present epoch, radar-ranging experiments (Reasonberg
     \etal 1979)
      imply $b>500$ or
      $\xi < .00025$. However, pure Brans-Dicke theory is not natural
      from a particle physics point-of-view.  For more plausible theories,
      such as scale-invariant
      models in which a dilaton acts as the Brans-Dicke field,
      the effective potential for $\phi$, which was taken to be
      zero in the example above,
      is expected to become non-zero when supersymmetry and/or
      conformal invariance is spontaneously broken at low energies  ---
      the dilaton develops a small mass (Green \etal 1987).
	 By this mechanism, the spontaneous symmetry breaking
 pins the value of
	 $\phi$. Once this occurs,
	  the theory reduces to conventional Einstein gravity
	   and, consequently,
	  the radar-ranging  experiments and other low-energy
	  tests place no constraint on $\xi$.  Provided the
	  inflation scale is higher than the symmetry breaking
	  scale,   the modifications of Einstein gravity can
  play their critical role in inflation without violating any
  of the conventional tests of general relativity.
	  Work on ``hyperextended'' inflation models went further to show that
	  virtually any $f(\phi)$ that increases monotonically with $\phi$
	  during inflation can be employed (Accetta and Steinhardt
	  1990).

The key conclusion is that there is a natural interplay between
inflation and gravity when virtually any modification is made to
Einstein gravity.  Inflation via an ordinary first order phase
triggers a change in the gravitational sector which has just
the desired  effect of turning inflation off! A beautiful conspiracy!
This natural  interplay
is, to this author, a compelling
reason to link inflation   to modified gravity theories.

\section{ Density perturbations in extended inflation and the
``big bubble'' problem}

Energy density fluctuations can come from two sources in
extended inflation, quantum fluctuations in the non-minimally
coupled field $\phi$ and from the collisions of bubbles of
true vacuum phase.
In this section,
 only the amplitude  of the fluctuation spectrum
 is discussed.  In Section~7, we will return to this
 subject to consider the deviation
 from strict scale-invariance.

Obtaining         small amplitude, adiabatic fluctuations represents a
severe constraint in new and chaotic inflation models (Steinhardt
and Turner 1984),
 but it  is a qualitatively different kind of constraint in
extended inflation.  In all inflationary
models, a nearly
scale-invariant spectrum of gaussian, adiabatic fluctuations is
generated due to de Sitter fluctuations in the fields evolving
continuously
during
inflation.  In new and chaotic inflation, the density fluctuation
amplitude can be shown to be:
\begin{equation}
\frac{\delta \rho}{\rho} \sim \frac{H^{2}}{\dot{\sigma}},
\end{equation}
where $\sigma$ is the scalar field which slowly rolls during the phase
transition and the expression is to be evaluated during the last 60
e-foldings of inflation when fluctuations on cosmological distance scales
were generated (Bardeen \etal 1983).
The amplitude is a ratio of two quantities that depend
upon the same field, $\sigma$:  $H^{2}$ depends upon the false vacuum
energy density of the $\sigma$ field and $\dot{\sigma}$ is the velocity
of the same field during inflation.  Hence, it is no surprise that the
dimensionful parameters cancel from the ratio and the amplitude reduces
to             a ratio of dimensionless parameters.  The isotropy of the
microwave background constrains  $\delta \rho / \rho < 10^{-5}$, which
imposes a constraint that the dimensionless parameters be unnaturally
small; for example, the quartic self-coupling of the $\sigma$ field is
constrained to be less than $10^{-12}$.

In extended inflation,
$\phi$, rather than $\sigma$,
evolves during inflation and      determines when
inflation ends.  Hence, one can show
that (La \etal 1990; Turner \etal 1990; Guth and Jain 1992):
\begin{equation}
\frac{\delta \rho}{\rho}= \frac{H^{2}}{\dot{\phi}}.
\end{equation}
In this case, $H^{2}$ depends upon the $\sigma$ field vacuum energy,
as before,
but the denominator depends on
$\phi$,          a completely independent field
characterized by
a different mass scale: $\phi \sim
{\cal O}(m_{Pl})$, where $m_{Pl}= 1.2 \times 10^{19}$~GeV is
the Planck mass.
Hence, $\delta \rho/\rho$ reduces to
a ratio
of dimensionful  quantities, $\delta \rho/\rho \propto
M^{2}/m_{Pl}^{2}$,
where $M$ is the energy scale for the phase transition.
By choosing $M < 10^{16}$ GeV or so, small amplitude fluctuations
can be obtained without fine-tuning of dimensionless couplings.
This relieves the usual fine-tuning problem;  and, since the field
can have strong couplings, reheating to high temperature after
inflation can be easily achieved.

A key constraint in extended inflation that has no analogue in new or
chaotic
inflation models is that there be an acceptable bubble-size
distribution (Weinberg 1990; La \etal 1990).
Producing too many bubbles of astrophysical size unacceptably
distorts the cosmic microwave background, creating a ``big
bubble problem.''
Initially, almost all the energy in the bubbles is contained in the
bubble-walls.  When the walls of different bubbles collide, the coherent
energy in the bubble walls is converted   into radiation and matter.
The radiation pressure drives the relativistic particles to stream
inwards and homogenize the energy over the bubble.  However, for very
large bubbles, there has not been sufficient time since inflation for
the radiation to reach the interior.  These bubbles represent large
amplitude ($\delta \rho/\rho \sim 1$) fluctuations with wavelength of
order the bubble radius.  One must ensure that the fractional volume
occupied by bubbles of order the horizon size at decoupling, $F(R>R_{D
})$, not exceed
$10^{-5}$ in order to avoid unacceptable distortions of the microwave
background.           Most bubbles are of order the horizon scale
at the end of  inflation, $H_{end}^{-1}$, where $H_{end}R_{D}
\approx 10^{25}$  ---     the ``dangerous,'' large bubbles lie
at the far tail of the bubble distribution.  Nevertheless, they
 provide a non-trivial constraint on inflationary models.

In old inflation,      the bubble nucleation parameter, $\epsilon =
\lambda/H^{4}$ is time-independent, so that the number of bubbles
produced per Hubble volume is the same each Hubble time.  Since
the bubble size depends on the age of the bubble, the bubble-size
distribution  is      scale-invariant: there are as many older
bubbles of  bigger radius as there are younger bubbles of
smaller radius.  In computing the fractional volume, $F(R>R_{D})$,
the bigger bubbles carry greater weight and $F(R>R_{D})$ is
divergent!  As Guth and Weinberg (1983) concluded, the bubbles produced
in old inflation produce an unacceptably inhomogeneous
universe.

In the original extended inflation model with non-minimal coupling,
$f(\phi) = \xi \phi^{2}$,  $H \propto 1/t$ or $\epsilon \propto t^{4}$.
Hence, with increasing      $\epsilon$, more younger/smaller bubbles
are produced compared to older/bigger bubbles.  One can show
that $F(R>R_{D})$ is no longer
 divergent (La \etal 1990; Weinberg 1990):
  $F(R>R_{D})
  =[H_{end}R_{D}]^{-32 \xi}$.   In order to satisfy  \\
    $F(R>R_{D}) <
    10^{-5}$, it is necessary that $\xi > .005$.  For a pure Brans-Dicke
    theory, this means that the Brans-Dicke parameter ($=1/8 \xi$) must be
    less than 25, which conflicts with well known radar-ranging
    limits.  However, as emphasized above, the radar-ranging constraints
    only apply for pure Brans-Dicke theory in which $\phi$ continues to
    evolve as  a free field through the present epoch.  If $\phi$ has a
    small potential, $W(\phi)$, which locks it to
     a constant value sometime
     after inflation, the radar-ranging limits are irrelevant and $\xi >.005$
     is acceptable (and even a natural range).

In hyperextended inflation (Accetta and Steinhardt 1990)
 and some generalized extended inflation
 models, $\epsilon$ increases exponentially fast near the end of
 inflation.  The ratio  of younger/smaller bubbles to older/bigger
 bubbles is greater still.  In this case, one finds $F(R>R_{D}) \approx
 {\cal O}(1)/[{\rm log}\;(H_{end}R_{D})]^3$, which generally satisfies
 the $10^{-5}$ bound without any special adjustment of parameters.
 Note, however, that the distribution is relatively flat so that
 there may be        a non-negligible number of bubbles extending
 up to moderate scales (few orders of magnitude smaller than $R_{D}$).

\section
{\bf  Gravitational radiation from bubble collisions}

The qualitative differences between extended inflation and prior
inflation models lead to some interesting, potentially different
predictions for the post-inflationary universe.

A fascinating suggestion is that
bubble nucleation will produce a
distinctive source of gravitational radiation (Turner and Wilczek
1991; Kosovsky \etal 1992).
This is a very narrow, intense peak in the gravitational wave
spectrum at short wavelengths (${\cal O}$(10~m)), which is
quite distinct from the broad gravitational wave background
produced by de Sitter fluctuations (Davis \etal 1992).
Crudely, a fraction of
the thermal radiation produced in bubble collisions will be in the form
of gravitational radiation with wavelength of order the horizon size at
the end of inflation red shifted to the present epoch.
The ratio of gravitational radiation to photons
will remain nearly constant until the present epoch, but the wavelength
will be redshifted.  Assuming that the critical temperature of the phase
transition is of order $10^{14}$ GeV or so  ($H \approx 10^{9}$~GeV),
produces a sharp peak
n the gravitational wave spectrum in the 10~m range with amplitude
roughly five orders of magnitude higher than the spectrum from
onventional inflationary models.  The amplitude and wavelength are in a
range that has an outside chance of
being detectable in forthcoming LIGO experiments.

The sharply peaked spectrum is not very sensitive to the bubble
distribution and, hence, is unaffected by how extended inflation
is completed.  The majority of the radiation comes from bubbles
with radii
of order  $H^{-1}$ where are copiously produced near the very
end of inflation.   Although Kosovsky \etal speculated that a similar
 peak in the spectrum might  be generated by a non-inflationary
 phase transition, more detailed (unpublished) work by this author
  has shown that the radiation from non-inflationary
  transitions is suppressed by a factor of $10^4$ or more.
  Briefly,
  the two ingredients necessary to produce a big peak are
  numerous bubbles with radius near $H^{-1}$ (the spectrum
  falls off sharply for smaller bubbles) and bubble walls in
  which the total energy density gained from tunneling is
  concentrated in a thin bubble wall.  These conditions are
  found only
  occur for bubbles created in a false vacuum background.
  Non-inflationary phase transitions
 do not produce bubbles of sufficient size (and the radiation is
 highly sensitive to the bubble size). Hence, it appears at present
 that
 a sharp, intense peak in the gravitational wave background is
 a unique signature of bubble collisions during an inflation.

Another by product of extended inflation is an easier relationship between
inflation and
cosmic string and global texture scenarios.  In
conventional inflation models, the couplings of the inflaton field are
fine-tuned to be so small that the reheating temperature is $10^{9}$ GeV
or less.  The reheat temperature is  so low that one would normally
conclude that it is problematic to proceed with   grand unification
scale baryosynthesis, or  to generate
cosmic string or
global texture defects that have been suggested as possible contributors
to large-scale galaxy structure.  In the past, additional fine-tuning is
assumed so that the massive defects can be generated at such a low
reheat temperature, and new methods for generating baryosynthesis are
invoked.
 I am told (by E. W. Kolb, private communication)
  that Zel'dovich once remarked that
  ``if you buy inflation, sell cosmic strings'' for this very reason.
  However,  with extended inflation, it is much easier to
  generate massive defects (Barrow \etal 1990).
    The most natural way is to choose the
    inflationary phase transition to be
     the same one that produces the desired
     defect.  Since the transition is completed by bubble nucleation, the
     usual Kibble mechanism (Kibble 1976)
      can be invoked to produce the defects.  (In new
      or chaotic inflation, the observable universe lies in one coherent
      region or one bubble, so no defects are generated.
       In extended
       inflation, the observable universe is the union of many bubbles, so
       many defects lie within the observable universe.)
       Hence, it is plausible to buy both inflation and cosmic strings (or
       global textures).

\section{A new mode of ending extended inflation}

In this section, we wish to examine carefully the difference
between  ``the end of inflation'' and ``the end of the
inflationary phase transition.''  The first refers to the
end of  superluminal expansion ($a(t)<t$) and the second
refers to the full conversion of false vacuum phase to true
vacuum phase.  In the past, the assumption has been that
the two  coincide: superluminal expansion ends when the
false vacuum decays.

However, it has recently been realized that  inflation can
end even while the universe is trapped in the false vacuum
phase. This new mode of ending inflation, which occurs for
a wide range of parameters, is attractive because it
automatically ensures an acceptable bubble distribution.

First,  consider the first
 mode of ending inflation in which inflation ends
only once the universe leaves the false vacuum.  It is
this mode that was
considered in the first extended inflation
papers  (La and Steinhardt 1989) and in all
analyses/criticisms  since (Weinberg 1990; La \etal 1990;
Turner \etal 1990).  This mode
 is exemplified by
 the original
Brans-Dicke theory with constant  Brans-Dicke
parameter $b$.  While the universe is trapped
in the false vacuum,
 the scale factor increases as a power-law (La and Steinhardt 1989)
 $a(t) \propto t^{b +
\frac{1}{2}}$, and the Hubble parameter,
$H = \dot{a}/a = (b+ \frac{1}{2})/t$,
decreases uniformly.  Hence, $\epsilon =\Gamma/H^4$ increases steadily, as
desired.

Density perturbations are generated during inflation from two sources.  Quantum
fluctuations in $\phi$ result in a nearly scale-invariant spectrum
(Turner \etal 1990; Guth and Jain 1992):
\begin{equation}
\frac{\delta \rho}{\rho}|_{H} \,(\lambda) \approx \frac{M_{F}^2}{m_{Pl}^2}
\;g(b)
\; \lambda^{4/(2b-1)},
\end{equation}
where $\frac{\delta \rho}{\rho}|_{H}(\lambda)$ is the perturbation amplitude as
length $\lambda$ re-enters the horizon after inflation;
$m_{Pl}^2 \equiv f(\phi)$ is the effective Planck mass at the end of
superluminal expansion; and $g(b)$ is an $b$-dependent
correction (Guth and Jain 1992)
that is close to unity except for $b \rightarrow 1$.
Note that a $\frac{\delta \rho}{\rho}|_{H}$ spectrum which scales as
$\lambda^m$
converts to a power spectrum $P(k) \propto k^{1-2m}$.
For $M_F/m_{Pl} \approx 10^{-(3-5)}$, but with no  fine-tuning of
couplings, $\frac{\delta \rho}{\rho}|_{H} (\lambda)$ is consistent with recent
COBE observations.
A
second  source of inhomogeneities comes from big bubbles
(Weinberg 1990; La \etal 1990). If inflation
continues until bubbles consume all remaining false vacuum,
the distribution of bubble sizes at the end of inflation is $
F(x>x_0) \approx \frac{1}{(1 + H_e x_0)^{4/b}}$,
 where $F(x>x_0)$ is the fractional volume occupied by bubbles of
 proper radius greater than $x_0$ and $H_{e}$ is the Hubble parameter
 at the end of inflation (La \etal 1990).
 To avoid  distortion of the CMB, we require $F(x>x_0)$ to be less than
$10^{-4}$ for
 bubbles of supercluster scale or larger, $H_{e} x_0 \ge 10^{25}$;  this
 implies $b < 25$.

 The serious criticism of this standard  mode  of ending inflation is
  the disappointingly limited
 range of workable models.
 Although the phase transition can be completed
 for any $b < \infty$, only $ b <25$  leads to
  acceptable inhomogeneities.  We shall see below that
  this implies a density perturbation power spectrum
  with $n\ltsim 0.84$.

The new mode ---  in which inflation ends  while the universe
remains  trapped in the false phase --- occurs if
$b$ is $\phi$-dependent and time-dependent (Crittenden and
Steinhardt 1992).
For
a very wide range of  polynomial or
exponential $f(\phi)$, $b$ decreases sharply with increasing $\phi$.
  When $\phi$ is small, $b >>1$ and
the universe expands superluminally.  Through
 the non-minimally coupling term, the false
vacuum energy density drives $\phi$ to steadily increase.
As $\phi$ grows, $b$ decreases and $a(t)$  accelerates
less rapidly.
 The boundary, $b = 1/2$, is critically important.
Recall that the scale factor grows as $a(t) \approx t^ {b + \frac{1}{2}}$,
for constant $b$ (La and Steinhardt 1989).
The expansion becomes subluminal ---
non-inflationary --- once $b$ drops below 1/2, even though
the universe remains
trapped in the false vacuum.  Even after this point,
$\phi$ continues to grow, the expansion rate
continues to decrease, and the bubble nucleation parameter, $\epsilon$,
continues to increase.  Some time after inflation has
stopped,  $\epsilon$ finally increases above unity and bubble nucleation ends
the transition to the false phase.  Figure 2 illustrates the typical
behavior for a wide range of $f(\phi)$ (e.g., $f(\phi) =
\phi^2 + \alpha \phi^4$).

\begin{figure}
\vspace*{3in}
\caption{  Typical behavior of
$b(\phi)= f/2[f']^2$ for polynomial or exponential
$f(\phi)$.  The new mode of
extended inflation  begins when $b \gg 1$
and ends when $b$ falls below 1/2. Bubble nucleation is negligible
until $b\ll 1/2$, too late for inflation to make many big bubbles.
}
\end{figure}

This new mode is important for several
reasons.  First, our recent
numerical calculations
suggest that viable models exist for a wide range of functional forms and
parameters, significantly wider than  the conventional mode.
(See Figs. 3 and 4 for some examples.)
Second, the big bubble problem is easily avoided because few bubbles
are nucleated ($\epsilon \ll 1$) while the universe is inflating.
(In the conventional mode, the troublesome CMB distortion comes from  inflation
of the many bubbles nucleated as $\epsilon$ gets close to unity
(La \etal 1990).)
 Almost the entire false vacuum is
converted to true vacuum by bubbles nucleated after the expansion has
become subluminal.  These uninflated bubbles   are
infinitesimally smaller than the horizon
at decoupling, much too small to distort the CMB.
Consequently, the big bubble problem is avoided altogether without
tuning of parameters.

\section{Scale-dependent density fluctuation spectrum}

We   next re-examine
the density perturbation spectrum generated
inflation to study the scale-dependence.  While it is often stated
that inflation predicts a ``scale-invariant'' spectrum
($\delta \rho/\rho$ in Eq.~(3) is constant), this is actually
only approximately correct.  In light of COBE DMR and forthcoming
experiments, a more precise analysis is essential.

  According to the recent COBE
results (Smoot \etal 1992),
the  CMB anisotropy can be
fit to a \underline{scale-free}
power spectrum, $P(k) \propto k^n$, where $n= 1.1 \pm 0.6$ ($1 \sigma$
limit).
Exponent
$n=1$ corresponds to an exactly
\underline{scale-invariant},
Harrison-Zeldovich spectrum.   New and chaotic
inflation
result in a nearly
 scale-invariant spectrum with logarithmic deviations
 ($\propto
{\rm log} \; \lambda$) (Bardeen \etal 1983).
This spectrum is determined by tracing Eq.~(3) during the
last 60 e-foldings of inflation using the slow-roll equation,
$3 H \dot{\sigma} = - V'(\sigma)$.
We have computed the exponent $n$ by fitting the density
perturbation spectrum to a scale-free form ($k^n$) fit
over astrophysical scales, $1-10^4$~Mpc.    For new or chaotic
inflation, our fit gives $n \approx 0.95.$
The full spectrum is essentially indistinguishable from
the case of extended
inflation where  $f(\phi)$ is exponential in $\phi$ (shown
in Figure~3 below).

In extended inflation with constant $b$ (e.g., Brans-Dicke model)
 the density perturbation spectrum  in
Eq.~(5) converts to a power spectrum with (Crittenden and Steinhardt
1992):
\begin{equation}\label{n}
n = \frac{2 b -9}{2 b-1}.
\end{equation}
For large $b$ (approaching the limit of
Einstein gravity), the spectrum is  near $n=1$.
As $b$ decreases, the
spectrum shifts to smaller values of $n$,
tipping the spectrum towards more power on large scales.

If extended inflation continues  until bubbles consume the
false phase altogether (the original mode considered by
La and Steinhardt (1989) and studied by later authors),
we   argued in Section~4
that $b$ must be less than 25 to avoid the big
bubble problem.
According to Eq.~(\ref{n}),  this requires $n<0.84$, a significant tilt away
from
scale-invariance.  This tilt corresponds to  increasing power on large
scales compared to scale-invariant spectra.  Some believe that this
much tilt leaves too little power on small scales to explain
galaxy formation, although this is tied to specific ideas (like
cold dark matter) about how fluctuations are processed once they
fall inside the horizon.  However, If the arguments are correct, it
is important to note that this would rule out this mode of
extended inflation altogether because of the big bubble problem.

Hence,
 a greater range of  $n$ is possible for the new mode of ending
 extending inflation.
In general, the amplitude $\frac{\delta \rho}{\rho}|_{H}(\lambda)$
is $\approx H^2/\dot{\phi}$
evaluated as scale $\lambda$ is stretched beyond the horizon during inflation.
Here,  $b$ (which
controls the rate of inflation) varies rapidly during inflation, especially
within the final e-foldings. In this case, there is no simple analytic
expression for the amplitude; rather, $\phi$, $b$, $H$, and
$\frac{\delta \rho}{\rho}|_{H}$ have to be tracked by numerically solving the
equations of
motion.
Our results are summarized in Figs.~3-5.
To compare with COBE (Smoot \etal 1992)
we  fit  the upper curves ($1-10^4$~Mpc)
to a scale-free
spectrum and determine $n$.  Fluctuations on these scales were
generated 50-60 e-foldings before the end of inflation and depend on
$b$ during that period (La \etal 1990; Turner \etal 1990;
Guth and Jain 1992).

\begin{figure}
\vspace*{6in}
\caption{$\delta \rho/\rho|_H$ as wavelength $\lambda$ enters the horizon
shown for
all observable scales (lower), with  blow-up showing large
scales  only (upper).  Scale-invariance or, equivalently,
a power spectrum with $n=1$, corresponds to a horizontal line.
 For  exponential
 $f(\phi)$, [in this example, $f(\phi)= M^2 {\rm exp} (.048 \phi/M)$],
 where
 $\phi \le M$ initially and $M\ge M_F$, the spectrum
 on large scales can be fit to $n\approx 1$ [in this
 example, $n=0.96$].
 }
 \end{figure}

 \begin{figure}
\vspace*{6in}
\caption{
 Same as Fig.~3 except for polynomial $f(\phi)$ [here, $f(\phi)=
 A (\phi^2 + B \phi^4/M^2)$, where $A=0.0061$ and $B=0.0021$].   The best-fit
to
 upper
 curve is $n=0.8$.
 Values $1.0 \ge n \ge 0.5$ can be obtained by varying $A$ and $B$, but
 $B/A$ must be increasingly tuned for yet smaller $n$.
 }
 \end{figure}

 \begin{figure}
 \vspace*{6in}
 \caption{
 Same as Fig.~3 except $f(\phi)= A (3.125 \, \delta \, \phi^2 -
 1.66  \phi^3/M^2 + 0.25 \phi^4)$
 is designed to have an inflection point  as $\delta \rightarrow 1$.
 Depending on parameters, a sharp dip may be created
  on astrophysical scales (lower and upper left; $A=0.2$ and
   $\delta=1.0001$);
   or,  it can be placed just beyond the horizon
     $\lambda \approx 10^4$~Mpc (upper right; $A= 0.059$ and
       $\delta=1.001$), which appears  scale-free within the
	 horizon with $n\approx 1.3$.
	 }
	 \end{figure}

We find three classes of behavior, depending on the functional
form for the non-minimal coupling, $f(\phi)$:
(1)~If $f(\phi)$ varies exponentially
with $\phi$ (see Fig.~3),  then $b$ remains very large until the
last 10 e-foldings of inflation.  Consequently, on scales $1-10^4$~Mpc,
the spectrum is flat, $n \rightarrow 1$, virtually indistinguishable from
new inflation. This regime was strictly disallowed in the original
 graceful exit mechanism for extended inflation.
  (2)~If $f(\phi)$ varies more slowly with $\phi$ (e.g.,
a polynomial), $b$ changes more gradually.  Hence, $b$ is
closer to 1/2  during the last 50-60 e-foldings of
inflation compared to (1), leading to a spectrum tipped towards $n<1$ on scales
$1-10^4$~Mpc.
(See Figure 4.)
However, $b$ cannot be too small and still satisfy all other
inflationary constraints (La \etal 1990).
We find that   $1 \gtsim n \gtsim 0.5$ is spanned by a plausible range
of parameters and  polynomial forms.
 (Yet smaller values of $n$ are possible, in principle, but only
if $f(\phi)$ is increasingly fine-tuned.)
(3)~Wild
variations  can be ``designed'' into the spectrum by special choices of
$f(\phi)$.
For example, $b(\phi) \equiv f/2[f']^2$ blows up at any near-inflection
point of $f(\phi)$. Consequently,
sharp ``dips'' in the perturbation spectrum
(see Fig.~5) are created as $\phi$ passes the near-inflection point.
The dips are similar to features created in ``designer
inflation'' models (Hodges \etal 1990;
Salopek \etal 1989) in new or chaotic inflation.
By choosing parameters carefully, the dip can be made to lie within
$1-10^4$~Mpc
scales (Fig.~5a), which is probably ruled out by COBE.   If the dip is placed
just beyond the
horizon, the spectrum on $1-10^4$~Mpc scales appears
to be scale-free (no dips), tipped towards $n>1$. Values
$1.3  \gtsim n \ge 1.0$ can be obtained.
See Fig.~ 5b.  What should be emphasized, though,
 is that
cases (1) and (2) occur for a robust set of functional forms and parameters,
whereas case (3) only occurs for a very special choice of $f(\phi)$.

In comparing our calculations to COBE, it should be emphasized that
we have computed  only the density
fluctuation spectrum.
Recently, it has been shown that gravity waves, produced during
inflation
at the same time as  the density perturbations, are
 the dominant source  of CMB distortions for
extended inflation
models with tilts $0.84\gtsim n \gtsim 0.5$ (Davis \etal 1992).
(In the cases with $n\ltsim 0.84$ considered in this paper,
e.g.,  Fig.~4,
we have been safe in ignoring the gravity waves  since
they do not change the spectral tilt  we
had been seeking to determine.) These new results are  expanded
upon in the paper by Smoot and Steinhardt that can be found
in this Proceedings.
There, it is shown that gravity waves for $n\ltsim0.84$
cannot be ignored if one tries to match COBE observations to measurements
on small ($<1^o$) angular scales and to models of galaxy formation.
Hence, COBE DMR and observations at small angular scales can be
used to determine the mode in which inflation was terminated.

\section{Conclusions}

Extended
inflation, based on the new mode in which expansion becomes
subluminal before escape from  the false vacuum, is   robust
and addresses many of the earlier criticisms of the
extended inflationary scenario.  Perhaps the most important
implications are a unique signature in the
gravitational radiation background on short wavelengths (${\cal O}$(10~m))
and
a revised prediction for the density fluctuation
spectrum.
With this mode,
  the  power spectrum for extended inflation is scale-free with a
wider range of possible power indices than previously thought possible:
$1.0 \ge n \gtsim 0.5$ for  plausible models, tilted towards
 more power on large scales ($n=1$ is exact Harrison-Zel'dovich).
  The range
 $n>1$ or strong deviations from a scale-free spectrum require unnatural
   functional forms and fine-tuning.  The predicted range is within the
  present $1\sigma$-limit for COBE (Smoot \etal 1992).  However, future
  experiments, including analysis of years two or more from COBE, should
  improve the measurements and provide useful new
  constraints on inflation models.

\section*{Acknowledgements}

The work reviewed in this paper includes research done in collaborations
between the author and many fine collaborators to whom
I am most grateful: R. Crittenden, F. Accetta,
D. La, M. S. Turner, A. Guth, G. Smoot, R. Davis, H. Hodges,
A. Albrecht, J. Bardeen, and E. Bertschinger.
I thank the organizers of the {\it Journ\'{e}es Relativistes 1992}
for inviting me to beautiful Amsterdam to attend the extremely
exciting meeting.  I especially thank R. Kerner,
S. Bais, W. van Leeuwen for   their
gracious hospitality and support in my behalf.
This work was supported
in part by
U.S. DOE Grant No. DOE-EY-76-C-02-3071.

\end{document}